\documentclass{elsarticle}

\usepackage{amsmath}
\usepackage{txfonts}
\usepackage{graphicx}
\usepackage{lipsum}
\usepackage{natbib}
\usepackage{xcolor}
\usepackage{color}
\usepackage{hyperref}
\usepackage{float}
\makeatletter
\def\ps@pprintTitle{%
 \let\@oddhead\@empty
 \let\@evenhead\@empty
 \def\@oddfoot{}%
 \let\@evenfoot\@oddfoot}
\makeatother

\biboptions{longnamesfirst,comma}
\begin{document}
\title{Quantum Information with Even/Odd States of Orbital Angular Momentum of Light}
\author[prl]{Chithrabhanu Perumangatt}\corref{cor}
\ead{chithrabhanu@prl.res.in}
\author[prl,iitgn]{Nijil Lal}   \author[prl]{ Ali Anwar} \author[uec]{ Salla Gangi Reddy} \author[prl]{
R. P. Singh}
\address[prl]{Physical Research laboratory, Navarangpura, Ahmedabad, India - 380009}
\address[iitgn] { IIT Gandhinagar, Palaj, Ahmedabad, India - 382355}
\address[uec]{University of Electro-communication, Chofu, Tokyo, Japan - 1828585}
\cortext[cor]{Corresponding author}

\begin{abstract}
We address the possibility of using even/odd states of orbital angular momentum (OAM) of photons for the quantum information tasks. Single photon qubit states and two photon entangled states in even/odd basis of OAM are considered. We present a method for the tomography and general projective measurement in even/odd basis. With the general projective measurement, we show the Bell violation and quantum cryptography with Bell's inequality. We also describe hyper and hybrid entanglement of even/odd OAM states along with polarization, which can be applied in the implementation of quantum protocols like super dense coding. 
\end{abstract}

\maketitle

\section{Introduction}

Quantum information protocols mainly rely on the fact that particle can be in a complex superposition of states. Polarization state of  photons is used extensively to implement many quantum protocols. The polarization of a photon spans in a two dimensional Hilbert space. So the polarization state of a photon is considered as a qubit. Also, one can generate photons entangled in polarization using spontaneous parametric down conversion (SPDC) of a laser beam. All four maximally entangled states, Bell states, can be achieved in the polarization degree of freedom (DOF). 

Orbital angular momentum (OAM) is another degree of freedom of photon that can be used in quantum protocols along with polarization so that the information carried per photon can be increased \cite{torres}.  OAM entanglement can also be achieved by SPDC and many quantum protocols were demonstrated using the same \cite{Mair2001,PhysRevLett.91.227902, PhysRevA.72.052114, PhysRevLett.89.240401,PhysRevA.67.052313, Dada2011, Malik2016, PhysRevA.88.032305}. The basis states of OAM span an infinite dimensional Hilbert space. This higher dimensionality is really useful for the denser coding of information in single photons \cite{PhysRevA.86.032334}. One can achieve OAM entanglement in higher dimensions which can be used for many quantum protocols. However, we often need to use two dimensional OAM states for the ease of measurements. Also for many protocols using hybrid-entanglement, entanglement between polarization and OAM of photons, we need the 2 dimensional OAM sub-space\cite{Nagali:10, PhysRevA.80.042322}. 

Experimentally the restriction of OAM states to 2D is done by post selection using diffractive holograms and a single mode fibre \cite{Mair2001}. This results in the loss of photons which reduces efficiency of the protocol. We investigate the possibility of using a 2D OAM space without any photon loss. This is possible since any infinite set of integers can be grouped into two natural categories: even and odd. In the case of OAM, this becomes possible because of an effective even/odd OAM sorter, an optical set up designed to separate even and odd states of OAM. However, the even/odd states of OAM have not been extensively explored for quantum information tasks. For that, we need to develop projective measurement in even/odd basis. We propose simple interferometric method for the projective measurements. 

We demonstrate the tomography of the even/odd states with projective measurement in Pauli's operator bases. We also describe hyper-entanglement and hybrid-entanglement with polarization as another DOF and propose interferometric set up for the spin orbit Bell state analysis (SOBA). Measurements for checking the Bell's inequality in even/odd OAM entanglement is discussed for the first time. This can be applied in entanglement based cryptographic protocol. It is theoretically impossible to distinguish all Bell states using local operations and classical communications (LOCC) \cite{Sibasish}. However, with hyper-entanglement and SOBA, one can distinguish all the Bell states of polarization using LOCC. Using the same, we describe efficient super dense coding.

\section{From Infinite Dimensional OAM Space to Two Dimensional Even/odd OAM Space}

The general infinite dimensional OAM space is spanned by the OAM values from $-\infty, .. -1,0,+1,... +\infty$. A general state in this infinite dimensional basis can be written as

\begin{equation}
\vert\psi\rangle = \sum_{m=-\infty}^{+\infty} c_{m} \vert m\rangle \label{1}
\end{equation}
with $\sum_{m=-\infty}^{+\infty}\vert c_m\vert^2 =1$. However, many quantum experiments were realized using the OAM qubits in the reduced Hilbert space \{$m,-m$\}. In such cases OAM encoding or measurements were performed using diffraction through holograms and the mode filtering. Basically here one neglects the photons generated with OAM $l\neq m,-m$ which result in photon loss. Moreover, the efficiency of mode filtering is also a limiting factor for quantum experiments with OAM. Thus to make an equivalent qubit state,  Eq. \ref{1} can be re written as 
\begin{equation}
\vert\psi\rangle = \sum\limits_k \left( c_{2k}\vert 2k\rangle + c_{2k+1}\vert 2k+1\rangle\right).\label{1a}
\end{equation}
We define the appropriate operators in order to  perform the measurements in the even/odd basis. The general projection operator is 
\begin{equation}
P(\theta,\phi) =\sum\limits_k \left(\text{cos}(\theta)\vert 2k\rangle+e^{i\phi}\text{sin}(\theta)\vert2k+1\rangle\right)\left(\text{cos}(\theta)\langle2k\vert+e^{-i\phi}\text{sin}(\theta)\langle 2k+1\vert\right).\label{1b}
\end{equation}
With these projective measurements we can consider the whole OAM state as a qubit state and use for quantum protocols. 
\\
 In the case of OAM entanglement, when we work with \{$m,-m$\} basis, photons corresponding to other modes are lost in the measurement. For example, when we pump a non-linear crystal for SPDC using a Gaussian beam, the signal and idler photons are entangled in OAM. The two photon state is given as 
\begin{equation}
 \vert\Psi\rangle_{12} =c_0 \vert 0\rangle\vert 0\rangle +  \sum_{m=1}^{+\infty} c_{m} \left( \vert m\rangle\vert -m\rangle+\vert -m\rangle\vert m\rangle \right)\label{2}
\end{equation}
with $\sum_{m=0}^{+\infty} \vert c_{m}\vert ^2 =1$
 In many of the OAM entanglement experiments, this state is projected in $\{+1,-1\}$ basis for treating it as a two qubit entangled state. In such cases, the probability of getting photons entangled in $\{+1,-1\}$ OAM states is $\vert c_1\vert ^2 \ll 1$. Thus most of the down converted photons remain unused. 

For even/odd OAM entanglement, we consider the parametric down conversion of an optical vortex of order 1 and having vertical polarization in a type I second order non-linear crystal. The state corresponding to the pair of photons produced by SPDC of this beam is given by
\begin{equation}
\vert\Psi\rangle_{12} = \sum_{m=-\infty}^{+\infty} c_m \vert m\rangle_1 \vert 1-m \rangle_2 \otimes \vert H\rangle_1\vert H\rangle_2\label{5}
\end{equation}  
By grouping all even and odd OAM states, one can rewrite the expression for the OAM state in Eq. \ref{5} as
\begin{equation}
\sum_{m=-\infty}^{+\infty} c_{m} (\vert m\rangle_1 \vert 1-m \rangle_2) \nonumber 
 =   \sum_{k=-\infty}^{+\infty} c_{2k} (\vert 2k\rangle_1 \vert 1-2k \rangle_2) + 
 \sum_{k=-\infty}^{+\infty} c_{1-2k} (\vert 1-2k\rangle_1 \vert 2k \rangle_2).\label{6}
\end{equation}
Thus
\begin{equation}
\vert\Psi\rangle_{12} =\left(\sum_{k=-\infty}^{+\infty} c_{2k} (\vert 2k\rangle_1 \vert 1-2k \rangle_2) + 
 \sum_{k=-\infty}^{+\infty} c_{1-2k} (\vert 1-2k\rangle_1 \vert 2k \rangle_2)\right)\otimes \vert H\rangle_1\vert H\rangle_2\label{6a}
\end{equation}  
From the conservation of OAM, we have 
\begin{equation}
\sum_{k=-\infty}^{+\infty} (c_{2k})^2 = \sum_{k=-\infty}^{+\infty} (c_{1-2k})^2= \frac{1}{2}\sum_{m=-\infty}^{+\infty} (c_{m})^2 = \frac{1}{2}.\label{6b}
\end{equation}
Thus one can arrive at an operational expression for even/odd OAM entanglement as
\begin{equation}
\vert\Psi\rangle_{12} = \frac{1}{\sqrt{2}}\left(\vert E\rangle_1 \vert O \rangle_2 +\vert O\rangle_1 \vert E \rangle_2 \right) \otimes \vert H\rangle_1\vert H\rangle_2. \label{7}
\end{equation}   
Here $\vert E\rangle$ and $\vert O\rangle$ correspond to the even/odd states on detection. Thus, we get a two qubit entanglement in OAM without loosing any photons.  
 
\section{State tomography for OAM states in even/odd basis}\label{sc.2}

 Similar to polarization, we need to find the Stokes vector for the super position state given in Eq. \ref{1a} by projective operators  which are
\begin{equation}
\begin{aligned}
   P_0 &=\sum_k (\vert 2k\rangle\langle 2k\vert+\vert 2k+1\rangle\langle 2k+1\vert)\\
   P_1&=\sum_k (\vert 2k\rangle\langle 2k\vert-\vert 2k+1\rangle\langle 2k+1\vert)\\ 
 P_2&=\sum_k (\vert 2k\rangle\langle 2k+1\vert+\vert 2k+1\rangle\langle 2k\vert)\\ 
 P_3&=\sum_k i(\vert 2k\rangle\langle 2k+1\vert-\vert 2k+1\rangle\langle 2k\vert)
\end{aligned}
\end{equation}
 Now, we define the Stokes parameters as
\begin{equation}
\begin{aligned}
 s_0 &= \langle \psi\vert P_0\vert\psi\rangle\equiv  \sum_k \left(c_{2k}c_{2k}^*+ c_{2k+1}c_{2k+1}^*\right) \\ 
 s_1 &=  \langle \psi\vert P_1\vert\psi\rangle \equiv  \sum_k \left(c_{2k}c_{2k}^*-c_{2k+1}c_{2k+1}^*\right)\\
 s_2 &=  \langle \psi\vert P_2\vert\psi\rangle \equiv  \sum_k \left(c_{2k}^* c_{2k+1}+ c_{2k}c_{2k+1}^*\right) \\
  s_3 &= \langle \psi\vert P_3\vert\psi\rangle\equiv  i\sum_k \left(c_{2k}^* c_{2k+1}- c_{2k}c_{2k+1}^*\right)
\end{aligned}
\end{equation}
\subsection{Measurements in Linear Even/odd Basis for $s_0$ and $s_1$}

\begin{figure}[h]
\centering
\includegraphics[scale=.5]{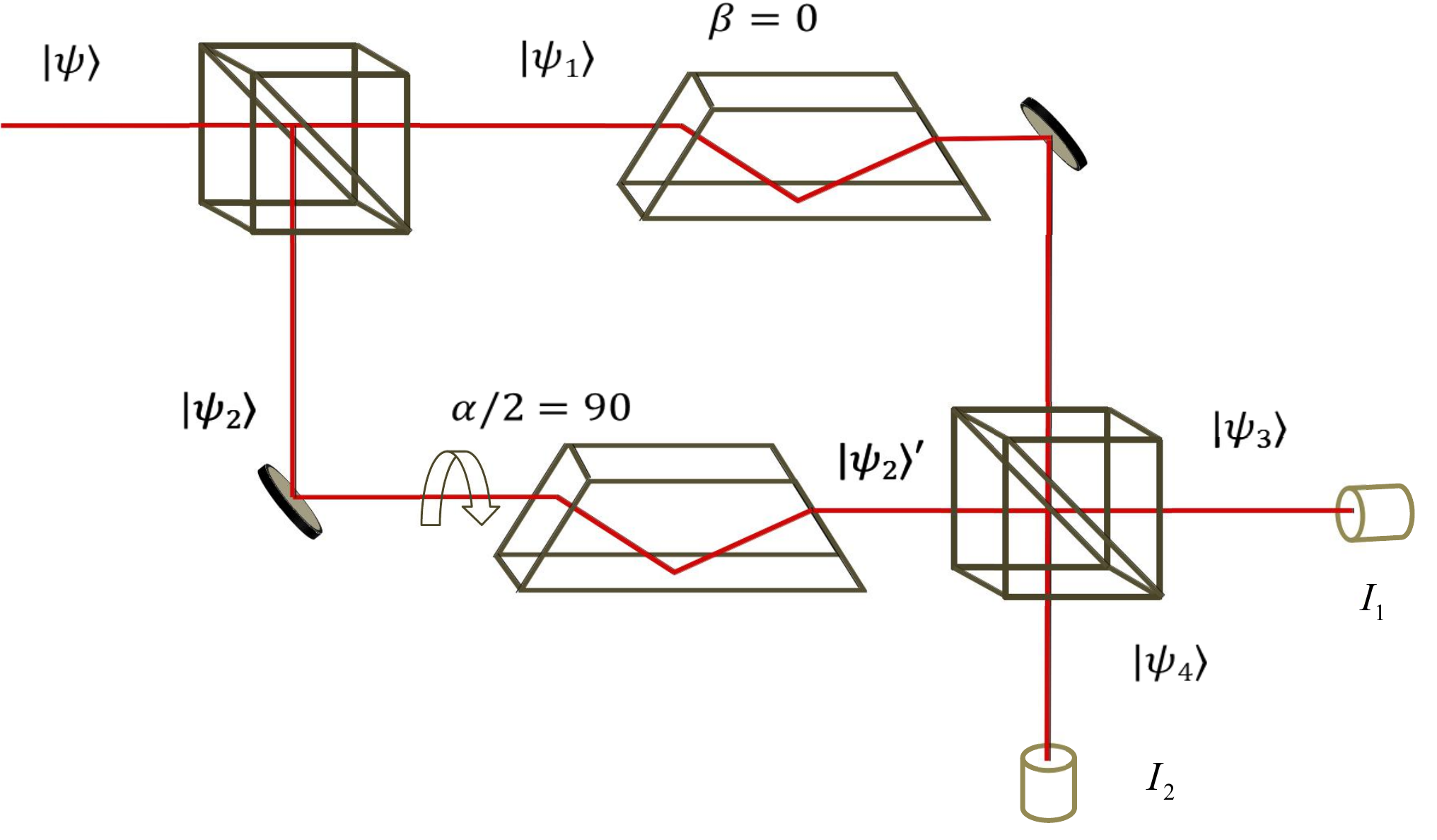}
 \caption{Even/odd OAM sorter}\label{fg.1}
\end{figure}

We consider an OAM sorter  \cite{Leach, PhysRevLett.92.013601, 2040-8986-18-5-054015} for the measurement of $s_0$ and $s_1$. The setup is given in Fig.~\ref{fg.1}
Consider a general even/odd OAM superposition state given in Eq.~\ref{1a}. Applying beam splitter operation, the state evolves through two arms of the interferometer with a $\frac{\pi}{2}$ phase. In the reflected arm, a dove prism is inserted which is rotated by an angle $ \frac{\alpha}{2}$. The  dove prism angle can be calibrated using an Hermite Gaussian beam $HG_{01}$ passing through it, as the rotation of the dove prism will result in rotation of the two lobes. Dove prism introduces an OAM dependent phase $\text{exp}(im\alpha)$, where $m$ correspond to OAM state $\vert m\rangle$. When $\alpha = \pi$, the even states will acquire a phase of $\text{exp}(i2k\pi)$ which leaves the state unchanged. However, odd states acquire a phase of $\text{exp}(i(2k+1)\pi)$ which brings a negative sign to all odd states. Thus the state $\vert\psi_2\rangle = \frac{i}{\sqrt{2}}\sum\limits_k \left( c_{2k}\vert 2k\rangle + c_{2k+1}\vert 2k+1\rangle\right)$ transforms to $\vert\psi_2\rangle' = \frac{i}{\sqrt{2}}\sum\limits_k \left( c_{2k}\vert 2k\rangle - c_{2k+1}\vert 2k+1\rangle\right)$. The state $\vert\psi_1\rangle = \frac{1}{\sqrt{2}}\sum\limits_k \left( c_{2k}\vert 2k\rangle + c_{2k+1}\vert 2k+1\rangle\right)$ of the other arm remains unchanged since the dove prism angle is $0^{\circ}$. These states combines at the second beam splitter. The phase due to reflections on both the beams are same and therefore neglected in the calculation. One port of the second beam splitter gives 
\begin{equation}
 \vert\psi_3\rangle = \frac{i}{2}\sum\limits_k \left( c_{2k}\vert 2k\rangle - c_{2k+1}\vert 2k+1\rangle\right)+\frac{i}{2}\sum\limits_k \left( c_{2k}\vert 2k\rangle + c_{2k+1}\vert 2k+1\rangle\right) = i\sum\limits_k  c_{2k}\vert 2k\rangle 
\end{equation}
Thus the detection (refer Fig.~\ref{fg.1}) yields 
\begin{equation}
I_1 = \sum\limits_k  \vert c_{2k}\vert^2
\end{equation}

 Similarly the other port gives
\begin{equation}
\begin{aligned}
 \vert\psi_4\rangle &= \frac{1}{2}\sum\limits_k \left( -c_{2k}\vert 2k\rangle + c_{2k+1}\vert 2k+1\rangle\right)+\frac{1}{2}\sum\limits_k \left( c_{2k}\vert 2k\rangle + c_{2k+1}\vert 2k+1\rangle\right)\\ &= \sum\limits_k  c_{2k+1}\vert 2k+1\rangle 
\end{aligned}
\end{equation}
 which gives $I_2 = \sum\limits_k  \vert c_{2k+1}\vert^2$ on detection (refer Fig.~\ref{fg.1}) . Thus we can calculate the stokes parameters $s_0= I_1+I_2$ and $s_1=I_1-I_2$ from this setup.

\subsection{Measurements in Diagonal Basis for the Estimation of  $s_2$}
\begin{figure}[h]
\begin{center}
\includegraphics[scale=.5]{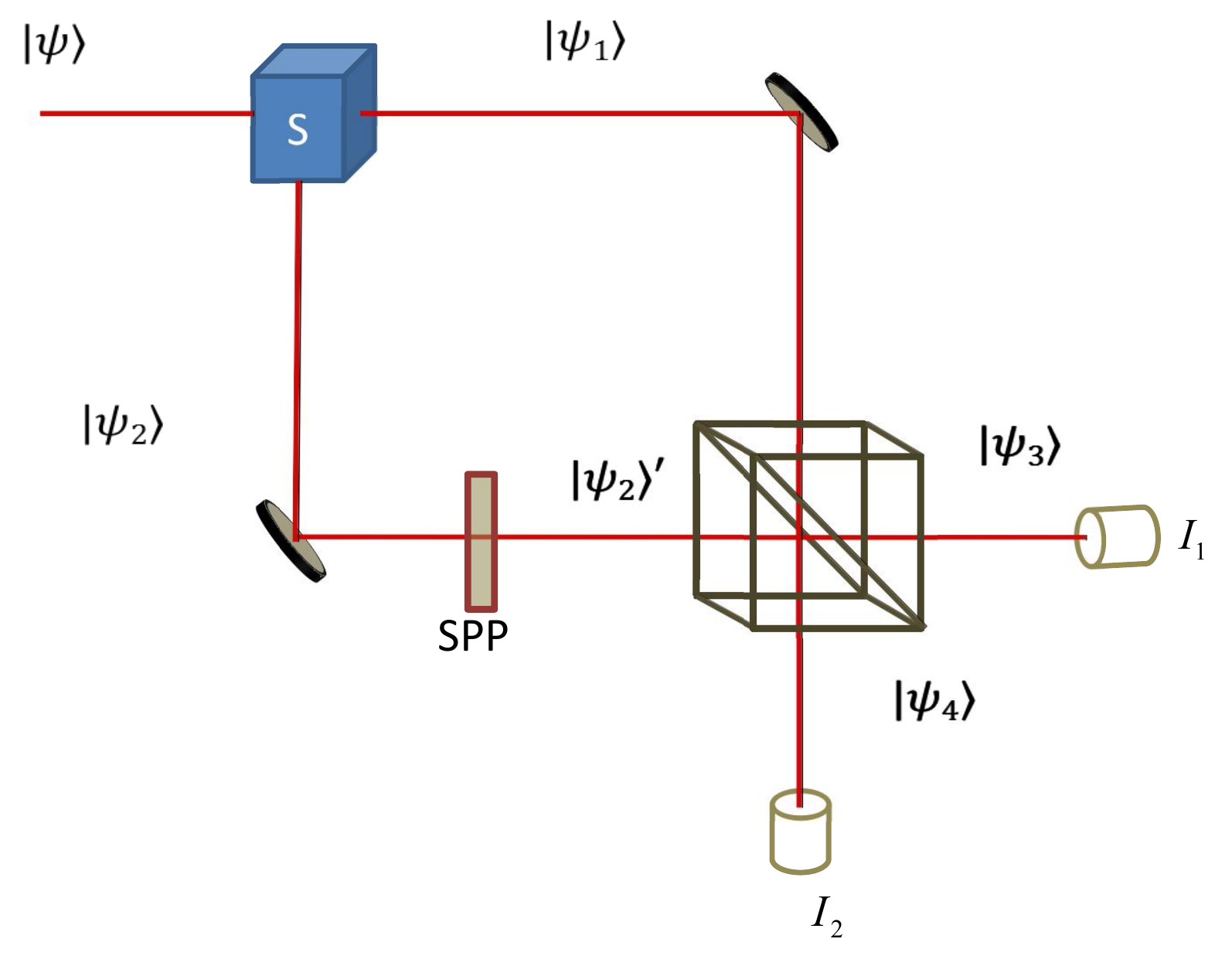}
 \caption{Setup for measuring stoke's parameter $s_2$}\label{fg.2}
\end{center}
\end{figure}

For measuring $s_2$, a modified Mach-Zhender  interferometer is introduced which contains a spiral phase plate in one arm and the first beam splitter is replaced by an OAM sorter. The setup is given in Fig.~\ref{fg.2}. As described earlier, a general OAM state $\sum\limits_k \left( c_{2k}\vert 2k\rangle + c_{2k+1}\vert 2k+1\rangle\right)$ is split into $i\sum\limits_k  c_{2k}\vert 2k\rangle $ and $  \sum\limits_k  c_{2k+1}\vert 2k+1\rangle $ in both ports. A spiral phase plate (SPP) of order $m=\pm1$ is introduced in one arm of the interferometer. This will act as a ladder operator in $\{..\vert -m\rangle ..,\vert -1\rangle,\vert 0\rangle,\vert 1\rangle,..,\vert m\rangle\} $ basis. However in the even/odd basis, SPP with $m=\pm1$ works as NOT gate. So the state $ \sum\limits_k  c_{2k+1}\vert 2k+1\rangle $  will convert to $  \sum\limits_k  c_{2k+1}\vert 2k\rangle $ . This is combined with state $i\sum\limits_k  c_{2k}\vert 2k\rangle $ on another 50:50 beam splitter. One port of the beam splitter yields the state 

\begin{equation}
\vert\psi_3\rangle = \frac{1}{\sqrt{2}}\sum\limits_k \left(  c_{2k+1}-c_{2k}\right)\vert 2k\rangle 
\end{equation}
On detection (refer Fig.~\ref{fg.2}) it gives
\begin{eqnarray}
\nonumber I_1 &=& \vert \frac{1}{\sqrt{2}}\sum\limits_k \left(  c_{2k+1}-c_{2k}\right)\vert 2k\rangle \vert ^2\\ &=& \nonumber \frac{1}{2}\left(\sum\limits_l \left(  c_{2l+1}^*-c_{2l}^*\right)\langle 2l\vert\right).\left( \sum\limits_k \left(  c_{2k+1}-c_{2k}\right)\vert 2k\rangle\right)  \\
\nonumber &=&\frac{1}{2} \sum\limits_{k,l}  \left(  c_{2l+1}^*-c_{2l}^*\right)\left(  c_{2k+1}-c_{2k}\right)\langle 2l\vert 2k\rangle\\
\nonumber &=& \frac{1}{2}\sum\limits_{k}  \left(  c_{2k+1}^*-c_{2k}^*\right)\left(  c_{2k+1}-c_{2k}\right) \\
I_1= \frac{1}{2}\sum\limits_{k} ( c_{2k+1}^*c_{2k+1}&-& c_{2k}^*c_{2k+1}-c_{2k}c_{2k+1}^*+c_{2k}^*c_{2k})
\end{eqnarray}
Similarly the other port of the BS (refer Fig.~\ref{fg.2}) gives
\begin{equation}
\vert\psi_4\rangle = \frac{i}{\sqrt{2}}\sum\limits_k \left(  c_{2k+1}+c_{2k}\right)\vert 2k\rangle 
\end{equation}
which gives
\begin{equation}
I_2= \frac{1}{2}\sum\limits_{k} (  c_{2k+1}^*c_{2k+1} +  c_{2k}^*c_{2k+1} +c_{2k}c_{2k+1}^*+c_{2k}^*c_{2k})
\end{equation}
Thus we obtain Stokes parameter $s_2$ by simply subtracting the intensities as 
\begin{equation}
I_2-I_1=\sum\limits_{k} ( c_{2k}c_{2k+1}^*+c_{2k}^*c_{2k+1}) = s_2. 
\end{equation}

\subsection{Measurement in Circular Basis for the Estimation of $s_3$}
\begin{figure}[h]
\begin{center}
\includegraphics[scale=.5]{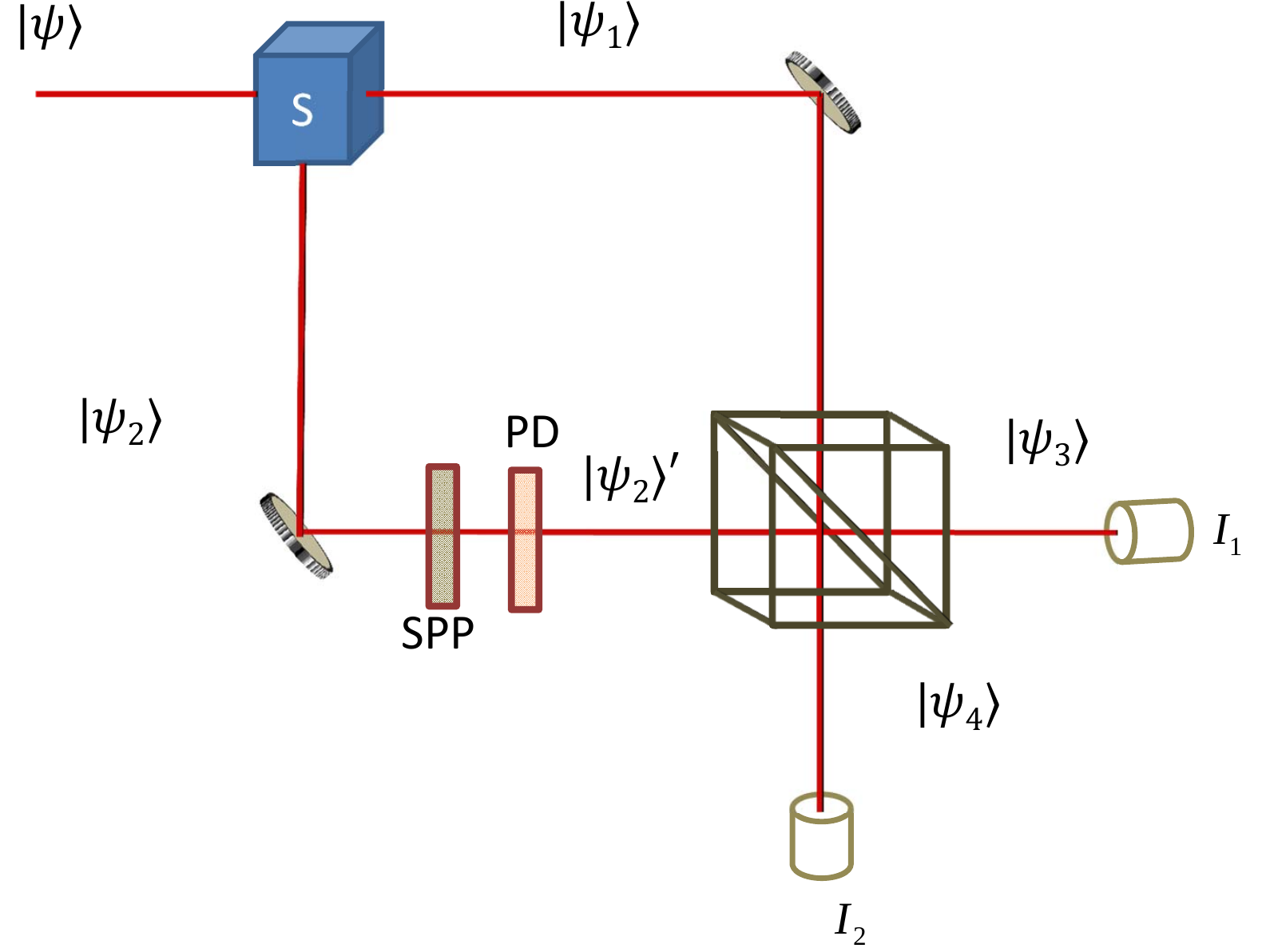}
 \caption{Setup for measuring stoke's parameter $s_3$. PD - phase delay}\label{fg.2a}
\end{center}
\end{figure}

For the measurement of $s_3$, an extra phase delay of $\text{exp}(i\pi/2) $ is inserted after the spiral phase plate as shown in Fig.~\ref{fg.2a}. Thus the states combining at the BS are  $  ic_2\vert E\rangle $ and  $  ic_1\vert E\rangle $. In this case one port of the BS (refer Fig.~\ref{fg.2a}) gives 
\begin{equation}
\vert\psi_3\vert = \frac{1}{\sqrt{2}}\sum\limits_k \left(  ic_{2k+1}-c_{2k}\right)\vert 2k\rangle
\end{equation}
so that
\begin{equation}
I_1=\frac{1}{2} \sum\limits_{k} (  c_{2k+1}^*c_{2k+1} - i c_{2k}^*c_{2k+1} +ic_{2k}c_{2k+1}^*+c_{2k}^*c_{2k})
\end{equation}

The other port of the BS (refer Fig.~\ref{fg.2a}) gives
\begin{equation}
\vert\psi_4\rangle = \frac{1}{\sqrt{2}}\sum\limits_k \left(  ic_{2k}-c_{2k+1}\right)\vert 2k\rangle 
\end{equation}
which gives
\begin{equation}
I_2= \frac{1}{2}\sum\limits_{k} (  c_{2k+1}^*c_{2k+1} + i c_{2k}^*c_{2k+1} -ic_{2k}c_{2k+1}^*+c_{2k}^*c_{2k})
\end{equation}
To obtain $s_3$ the intensities are subtracted as
\begin{equation}   
I_2-I_1=i\sum\limits_{k} ( c_{2k}c_{2k+1}^*-c_{2k}^*c_{2k+1}) = s_3. 
\end{equation}

Thus, once can do the complete state tomography in even/odd OAM states. 

\subsection{General Linear Basis Projection  }\label{sc.21}
A general linear projection is essential for the measurement of Bell's inequality and quantum cryptography. The general linear projections are given as
\begin{align}
\nonumber P_{\theta} &= \sum\limits_{k}(\text{cos}^2(\theta) \vert 2k \rangle\langle 2k\vert+ \text{sin}(\theta)\text{cos}(\theta)\vert 2k \rangle\langle 2k+1\vert+ \\ & \ \ \ \ \ \text{sin}(\theta)\text{cos}(\theta)\vert 2k+1 \rangle\langle 2k\vert+ \text{sin}^2(\theta) \vert 2k+1\rangle\langle 2k+1\vert ) \label{8}\\
\nonumber P_{\theta^{\perp}} &=\sum\limits_{k}(\text{cos}^2(\theta) \vert 2k \rangle\langle 2k\vert- \text{sin}(\theta)\text{cos}(\theta)\vert 2k \rangle\langle 2k+1\vert- \\ & \ \ \ \ \ \text{sin}(\theta)\text{cos}(\theta)\vert 2k+1 \rangle\langle 2k\vert+ \text{sin}^2(\theta) \vert 2k+1\rangle\langle 2k+1\vert ) \label{9}
\end{align}
These operators acting on $\vert \psi\rangle$ will give 
\begin{align}
C(\theta) &= \langle\psi\vert P_{\theta}\vert\psi\rangle = \sum\limits_{k}\vert (\text{cos}(\theta)c_{2k}+\text{sin}(\theta)c_{2k+1})\vert^2 \label{10}\\
C(\theta^\perp) &= \langle\psi\vert P_{\theta^{\perp}}\vert\psi\rangle = \sum\limits_{k} \vert(\text{sin}(\theta)c_{2k}-\text{cos}(\theta)c_{2k+1})\vert^2 \label{11 }
\end{align}
Consider the setup given in Fig. \ref{fg.2}. In the case of measurement of $s_2$ we used a 50:50 beam splitter. Now consider a beam splitter with transmission coefficient $\text{cos}(\theta)$ and reflection coefficient $\text{sin}(\theta)$ instead of 50:50 beam splitter.  Thus the states after the beam splitter becomes, 
\begin{align}
\psi_4 &=  \sum\limits_{k} (\text{cos}(\theta)c_{2k}+\text{sin}(\theta)c_{2k+1})\vert 2k\rangle   \\\
\psi_5  &= i\sum\limits_{k} (\text{sin}(\theta)c_{2k}-\text{cos}(\theta)c_{2k+1}\vert 2k\rangle 
\end{align}
On detection, we get 
\begin{align}
I_1 & = \sum\limits_{k}\vert (\text{cos}(\theta)c_{2k}+\text{sin}(\theta)c_{2k+1})\vert^2 \\
I_2 &=  \sum\limits_{k} \vert(\text{sin}(\theta)c_{2k}-\text{cos}(\theta)c_{2k+1})\vert^2 
\end{align}
Thus we achieve the general linear projections on a given state. 

Alternatively, one can also implement the same measurement with polarization as an additional degree of freedom.  We consider the initial photons with OAM state $\psi_1$ as horizontally polarized. In both arms of the interferometer, a half wave plates at angle $\theta/2$ is introduced which will convert the horizontal polarization to any other linear polarization along $\hat{\theta}$. The beam splitter is replaced by a polarizing beam splitter. It will transmit horizontal polarization and reflect vertical polarization. Thus by changing the HWP's angle $(\theta/2)$ we can tune the transmission and reflection function as $\text{sin}(\theta)$ and $ \text{cos}(\theta) $. The polarization assisted projection is easy to implement. However, when we consider OAM and polarization together for quantum protocols, for example hyper-entanglement, this cannot be used since the polarization operations affect the entanglement.

\section{Hyper-entanglement, hybrid entanglement and SOBA}
 Along with the OAM entanglement in even/odd states one can have polarization entanglement between the two photons \cite{PhysRevLett.95.260501}. For this, one needs to use a cascaded type I non linear crystals with optics axis perpendicular to each other for the parametric down conversion of a light beam of azimuthal index 1. The generated state will be
 
 \begin{equation}
 \vert\psi\rangle_{12} = \frac{1}{2}\left(\vert H\rangle_{1} \vert H \rangle_{2}+\vert V\rangle_{1} \vert V\rangle_{2} \right)\otimes\left(\vert E \rangle_{1} \vert O\rangle_{2}+\vert O \rangle_{1} \vert E\rangle_{2}\right)
\end{equation}  
This state has many applications including hyper entangled assisted Bell state analysis (HBSA) and super dense coding. Note that in hyper-entanglement, the polarization and OAM states are always separable. In other words, there is no entanglement between polarization and OAM. 

\begin{figure}
\centering
\includegraphics[scale=.5]{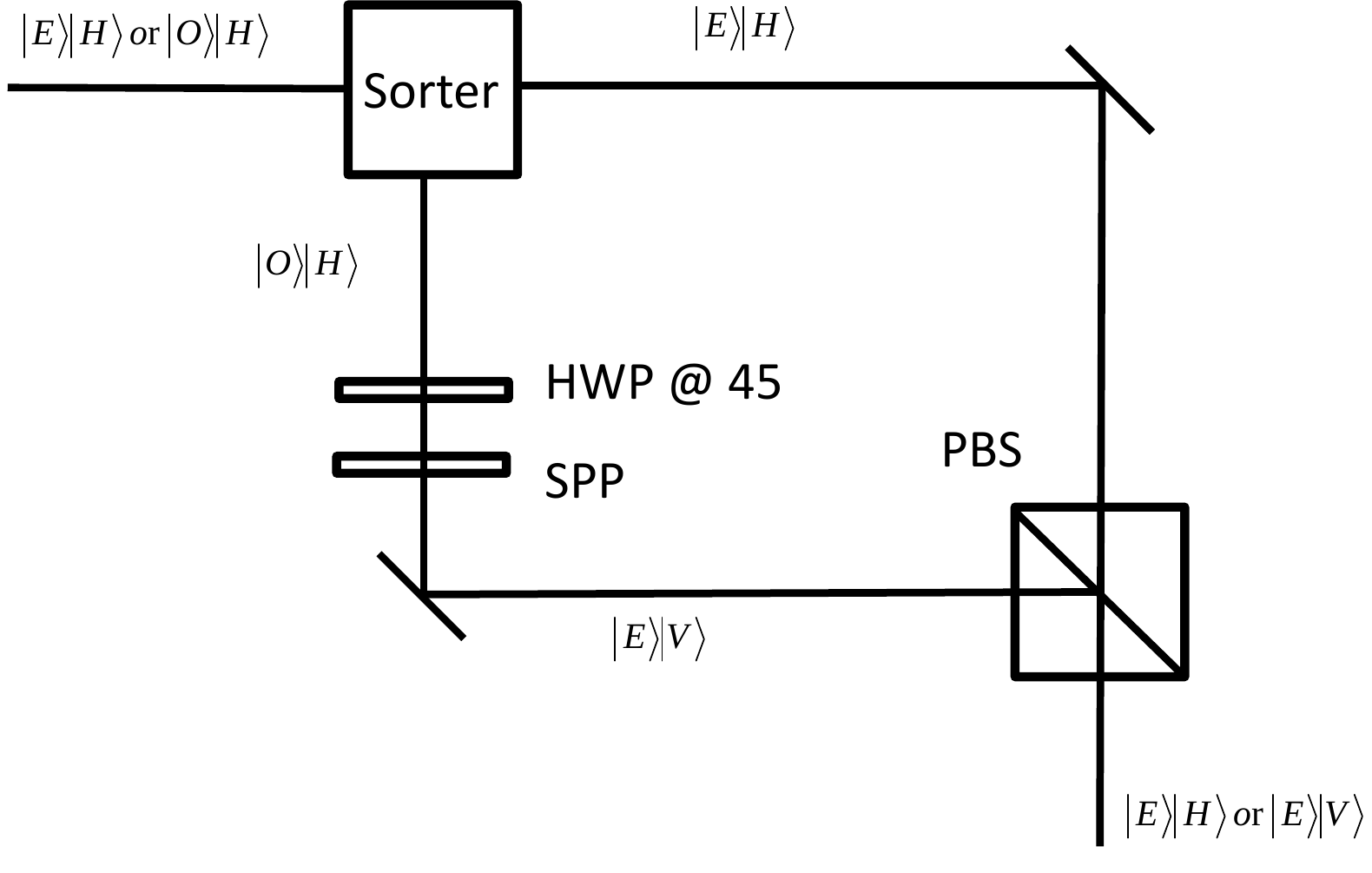}\label{fg.3}
\caption{Modified OAM controlled polarization $C_{NOT}$ gate. }
\end{figure}

Hybrid entanglement, as the name suggests, is the entanglement between two independent properties of light. The state of a single particle or two particles in two degrees of freedom are non-separable in the case of hybrid entanglement. In the biphoton systems, the polarization of one photon and orbital angular momentum of other, can be made non-separable. To generate this state, we consider two photons entangled in OAM even/odd states but separable in polarization, Eq. \ref{7}. We consider a  modified OAM controlled polarization $C_{NOT}$ gate ($^oC_p$) \cite{perumangatt2} with an extra spiral phase plate as given in Fig. \ref{fg.3}. This $C_{NOT}$ gate will do a NOT operation on both control and target. 
This modified $^oC_p$ is introduced to the first photon of the state, given in Eq. \ref{5}. Thus the two photon state will become
\begin{equation}
\vert\psi\rangle_{HE} = \frac{1}{\sqrt{2}} \left(\vert H\rangle_1\vert O\rangle_2+\vert V\rangle_1\vert E\rangle_2\right)\vert E\rangle_1\vert H\rangle_2
\end{equation}
This is an interesting case, since there is no OAM-OAM entanglement or polarization polarization entanglement between the photons. However, the polarization state of photon 1 and OAM state of photon 2 are non-separable. With this state one can steer the OAM state of photon 2 by polarization measurements in photon 1. 

\begin{figure}
\centering
\includegraphics[scale=.5]{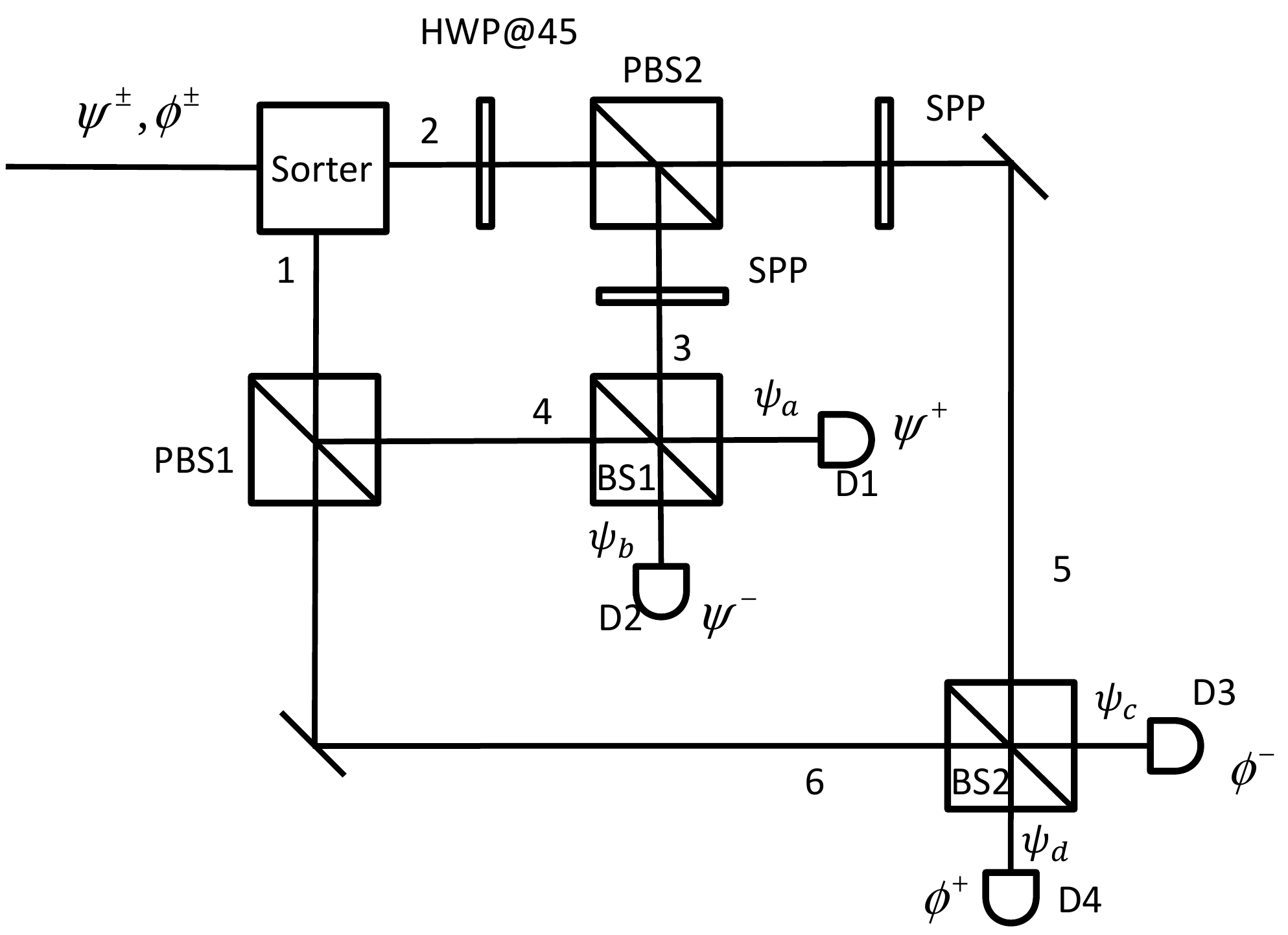}
\caption{Set up for spin orbit Bell state analysis \label{fg.4} }
\end{figure}

Single photon non-separable state also considered as a hybrid entangled state is the one in which the OAM and polarization of a single photon are inseparable. However, this won't give rise to any non-local effects and hence there are objections to call such states as entangled. We consider the state given in Eq. \ref{5} where state is post selected to state $\vert O\rangle_2\vert H\rangle_2$ 

\begin{equation}
\vert\psi\rangle'_1=\vert E\rangle_1\vert H\rangle_1.
\end{equation}  
Here $\vert\psi\rangle'$ corresponds to the state of photon 1 upon the post selection of photon 2 to the state $\vert O\rangle_2\vert H\rangle_2$.  Now, we apply a Hadamard operation in polarization using a half wave plate at $22.5^{\circ}$ and a polarization controlled OAM $C_{NOT}$ gate $^pC_o$, we obtain
\begin{equation}
\vert\psi\rangle'_1=\frac{1}{\sqrt{2}}\left(\vert E\rangle_1\vert H\rangle_1+\vert O\rangle_1\vert V\rangle_1\right).
\end{equation}  
Here the polarization and OAM of a single photon are non-separable. One can construct a complete Bell basis as
\begin{equation}
\begin{aligned}
\psi^{+} &= \frac{1}{\sqrt{2}}\left(\vert E\rangle_1\vert H\rangle_1+\vert O\rangle_1\vert V\rangle_1\right)\label{17} \\
\psi^{-} &= \frac{1}{\sqrt{2}}\left(\vert E\rangle_1\vert H\rangle_1-\vert O\rangle_1\vert V\rangle_1\right) \\
\phi^{+} &= \frac{1}{\sqrt{2}}\left(\vert O\rangle_1\vert H\rangle_1+\vert E\rangle_1\vert V\rangle_1\right) \\
\phi^{-} &= \frac{1}{\sqrt{2}}\left(\vert O\rangle_1\vert H\rangle_1-\vert E\rangle_1\vert V\rangle_1\right)
\end{aligned}
\end{equation}
A spin orbit Bell state analyser (SOBA) is a set up which distinguishes all the single photon spin orbit Bell states given above. Fig. \ref{fg.4} describes the proposed setup for the SOBA. Consider the above four Bell states as inputs of the SOBA set up. Initially an even/odd sorter sorts according to the OAM state. Lets first consider $\vert\psi^{\pm}\rangle$ as the input states. The 
port 1 \& 2 will be 

\begin{equation}
\begin{aligned}
\text{port 1} \rightarrow \pm \frac{1}{\sqrt{2}}  \vert O\rangle_1\vert V\rangle_1, \
\text{port 2} \rightarrow  \frac{i}{\sqrt{2}}  \vert E\rangle_1\vert H\rangle_1
\end{aligned}
\end{equation} 
A half wave plate is introduced in port 2 which will convert $\vert H\rangle_1$ to $\vert V\rangle_1$ which will be the input of PBS 2. The PBS 2 will reflect the state since it is vertically polarized. In the reflected port a spiral phase plate of order 1 is introduced which will convert $\vert E\rangle_1$ to $\vert O\rangle_1$. At the same time, PBS 1 reflects the state in port 1.  Thus at input ports 3 and 4 of the BS 1 we get
\begin{equation}
\begin{aligned}
\text{port 3} \rightarrow  \frac{i}{\sqrt{2}}&  \vert O\rangle_1\vert V\rangle_1\\
\text{port 4} \rightarrow \mp \frac{1}{\sqrt{2}} & \vert O\rangle_1\vert V\rangle_1
\end{aligned}
\end{equation} 
Two output ports of the BS 1 gives 
\begin{equation}
\begin{aligned}
\psi_a &= -\frac{1}{2}(\vert O\rangle_1\vert V\rangle_1 \pm \vert O\rangle_1\vert V\rangle_1)\\
\psi_b &= i\frac{1}{2}(\vert O\rangle_1\vert V\rangle_1 \mp \vert O\rangle_1\vert V\rangle_1)
\end{aligned}
\end{equation}
So $\vert\psi^{+}\rangle$ will go to detector $D_1$ and $\vert\psi^{-}\rangle$ will go to detector $D_2$. 

Now if the input states are $\vert\phi^{\pm}\rangle$ ports 1 and 2  will have states 
 \begin{equation}
\begin{aligned}
\text{port 1} \rightarrow  \frac{1}{\sqrt{2}}  \vert O\rangle_1\vert H\rangle_1
\text{port 2} \rightarrow \pm \frac{i}{\sqrt{2}}  \vert E\rangle_1\vert V\rangle_1
\end{aligned}
\end{equation} 
In port 2 after the HWP the polarization will convert to $\vert H\rangle_1$, and in both ports the state will be transmitted by the PBS1 and PBS2. In port 2, after PBS2, a SPP is inserted. Hence at the input ports 5 and 6 of the BS 2 the states will be    
\begin{equation}
\begin{aligned}
\text{port 5} \rightarrow \pm \frac{i}{\sqrt{2}}&  \vert E\rangle_1\vert H\rangle_1\\
\text{port 6} \rightarrow  \frac{1}{\sqrt{2}} & \vert E\rangle_1\vert H\rangle_1
\end{aligned}
\end{equation} 
Now two output ports of the BS 2 gives 
\begin{equation}
\begin{aligned}
\psi_c &= -\frac{1}{2}(\vert E\rangle_1\vert H\rangle_1 \mp \vert E\rangle_1\vert H\rangle_1)\\
\psi_d &= \frac{1}{2}(\vert E\rangle_1\vert H\rangle_1 \pm \vert E\rangle_1\vert H\rangle_1)
\end{aligned}
\end{equation}
So $\vert\phi^{-}\rangle$ will go to detector $D_3$ and $\vert\phi^{+}\rangle$ will go to detector $D_4$. 
Hence we can have complete deterministic unambiguous Bell state analysis. 

\section{Quantum Information Protocols Using Even/odd OAM States}

Our final aim is to implement quantum information protocols using even/odd state of OAM. This could increase the efficiency of quantum communication with photons since we can have a high bright source of OAM entanglement equivalent to the polarization entanglement. We show that the even/odd entangled states violate Bell's inequality which has direct application in Ekert protocol. With hyper-entanglement we show it can be used in superdense coding. 

\subsection{Violation of Bell's Inequality and Ekert Protocol}

As explained in section \ref{sc.21}, one can do projections to general linear states. Now consider a two photon state entangled in even/odd OAM state produced by the parametric down conversion of an optical vortex of order one \cite{PhysRevA.69.023811, Khoury2011}. The state is given as

\begin{equation}
\vert\Psi\rangle_{12} =\left(\sum_{k=-\infty}^{+\infty} c_{2k} (\vert 2k\rangle_1 \vert 1-2k \rangle_2) + 
 \sum_{k=-\infty}^{+\infty} c_{1-2k} (\vert 1-2k\rangle_1 \vert 2k \rangle_2)\right)\otimes \vert H\rangle_1\vert H\rangle_2
\end{equation}  

\begin{figure}[h]
\centering
\includegraphics[scale=.4]{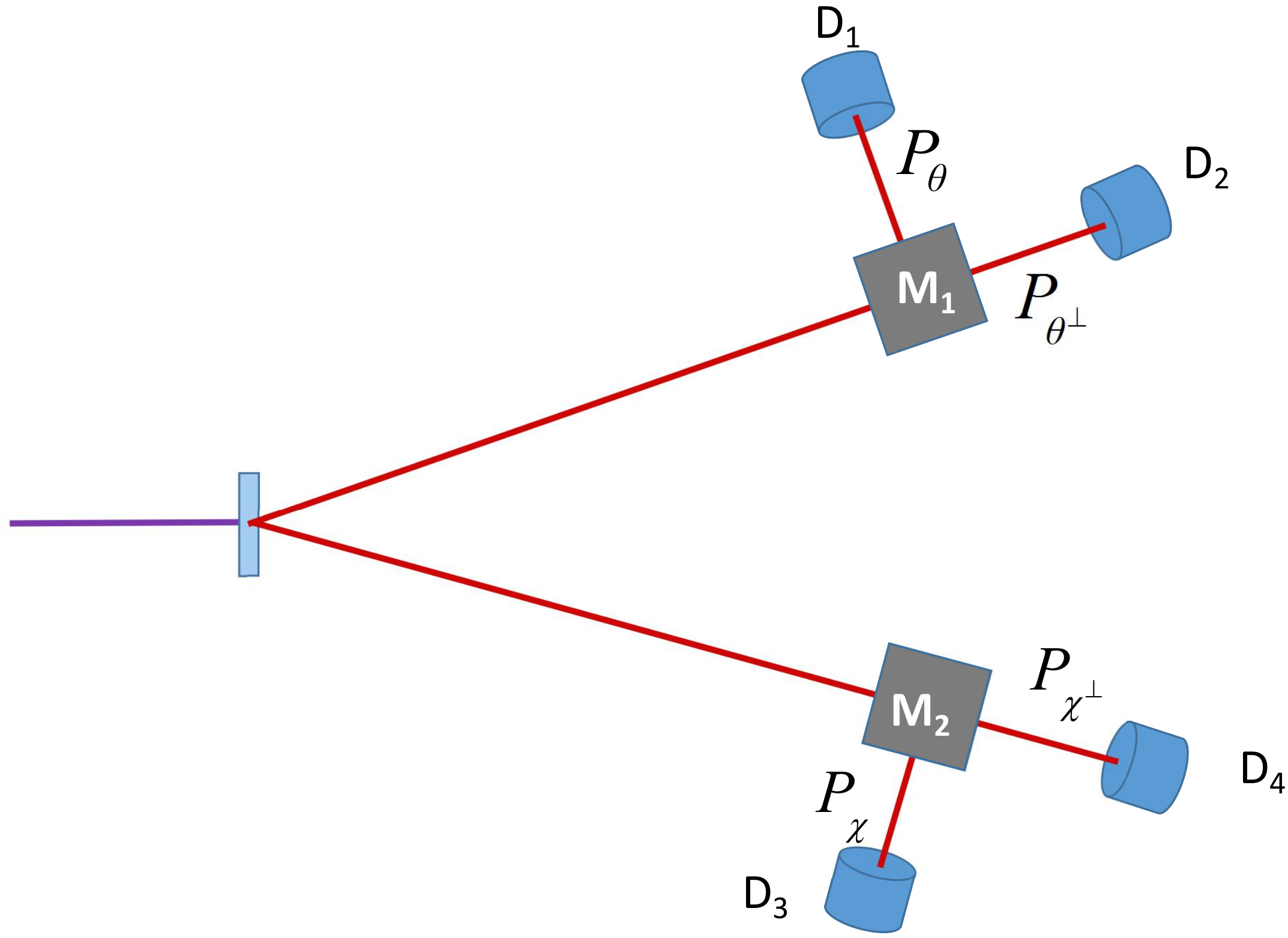}
 \caption{Setup for checking Bell's inequality and quantum cryptography. M1 and M2 are two measurements explained in Section \ref{sc.21} with angles $\theta$ and $\chi$ respectively. }\label{fg.5}
\end{figure}

 Consider the photon 1 with Alice and photon 2 with Bob. Alice does $P_{\theta}$ and $ P_{\theta^\perp}$ measurements on her photon and Bob does $P_{\chi}$ and $ P_{\chi^\perp}$ measurements on his photon. Eq. \ref{8} and \ref{9} describes $\vert\theta\rangle, \vert\theta^\perp\rangle$ and $\vert\chi\rangle, \vert\chi^\perp\rangle$ with replacing $\theta$ by $\chi$. The measurement is given in section \ref{sc.21}, $\theta$ and $\chi$ correspond to $\text{sin}^{-1}(t)$ where $ t$ is the transmissivity of the beam splitter. In the case of polarization assisted projection as given in Fig. \ref{fg.5},  $\theta$ and $\chi$ correspond to twice of the angle of the HWPs . 

Instead of intensity/photon counting output, which is described in \ref{sc.21}, here we consider the coincidence between Alice's and Bob's detectors. 

Coincidence between $D_1$ and $D_3$ will give the measurement result $_{12}\langle\Psi\vert P_{\theta}\otimes P_{\chi}\vert\Psi\rangle_{12}$, where $P_{\theta}$ and $P_{\chi}$ are defined by Eq. \ref{9} with angle $\theta$ and $\chi$ that  act on photon 1 and 2 respectively.

\begin{equation}
D_{13} = c(\theta,\chi) = {}_{12}\langle\Psi\vert P_{\theta}\otimes P_{\chi}\vert\Psi\rangle_{12}
\end{equation}
Similarly 
\begin{align}
D_{14} =& c(\theta,\chi_{\perp}) = _{12}\langle\Psi\vert P_{\theta}\otimes P_{\chi^{\perp}}\vert\Psi\rangle_{12}\\
D_{23} =& c(\theta_{\perp},\chi) =  _{12}\langle\Psi\vert P_{\theta^{\perp}}\otimes P_{\chi}\vert\Psi\rangle_{12}\\
D_{23} =& c(\theta_{\perp},\chi_{\perp}) = _{12}\langle\Psi\vert P_{\theta^{\perp}}\otimes P_{\chi^{\perp}}\vert\Psi\rangle_{12}
\end{align}
The operator is defined as 
\begin{align}
\nonumber P_{\theta}\otimes P_{\chi} &\equiv \sum\limits_{k}(\text{cos}^2(\theta) \vert 2k \rangle\langle 2k\vert+ \text{sin}(\theta)\text{cos}(\theta)\vert 2k \rangle\langle 1-2k\vert+ \\ &\nonumber \ \ \ \ \ \text{sin}(\theta)\text{cos}(\theta)\vert 1-2k \rangle\langle 2k\vert + \text{sin}^2(\theta) \vert 1-2k\rangle\langle 1-2k\vert )\otimes  \\ \nonumber & \sum\limits_{l}(\text{cos}^2(\chi) \vert 2l \rangle\langle 2l\vert+ \text{sin}(\chi)\text{cos}(\chi)\vert 2l \rangle\langle 1-2l\vert+ \\ & \ \ \ \ \ \text{sin}(\chi)\text{cos}(\chi)\vert 1-2l \rangle\langle 2l\vert+ \text{sin}^2(\chi) \vert 1-2l\rangle\langle 1-2l\vert )
\end{align}

Operating this on Eq. \ref{6a}
\begin{align}
\nonumber {}_{12}\langle \Psi\vert P_{\theta}\otimes P_{\chi}\vert\psi\rangle_{12} = & \frac{1}{2} \left(\sum_{m=-\infty}^{+\infty} c_{2m}^* (_1\langle 2m\vert _2\langle 1-2m \vert) + 
 \sum_{m=-\infty}^{+\infty} c_{1-2m}^* (_1\langle 1-2m\vert _2\langle  2m \vert)\right)\\ & P_{\theta}\otimes P_{\chi} \left(\sum_{n=-\infty}^{+\infty} c_{2n} (\vert 2n\rangle_1 \vert 1-2n \rangle_2) + 
 \sum_{n=-\infty}^{+\infty} c_{1-2n} (\vert 1-2n\rangle_1 \vert 2n \rangle_2)\right)
\end{align}

gives 
\begin{equation}
\begin{aligned}
C(\theta,\chi) =  \sum_{m,n,k,l= -\infty}^{+\infty}\bigg [ &  \text{cos}^2(\theta)\text{sin}^2(\chi)c_{2m}^*c_{1-2n}\langle 2m\vert 2k\rangle  \langle 2k\vert 2n\rangle \langle 1-2m\vert 1-2l\rangle \\ &  \langle 1-2l\vert 1-2n\rangle+ \text{cos}(\theta)\text{sin}(\theta)\text{cos}(\chi)\text{sin}(\chi)c_{2m}^*c_{2n}  \langle 2m\vert 2k\rangle \\ & \langle 1-2k\vert 1-2n\rangle   \langle 1-2m\vert 1-2l\rangle\langle 2l\vert 2n\rangle+
\text{cos}(\theta)\text{sin}(\theta)\\ & \text{cos}(\chi)\text{sin}(\chi)c_{1-2m}^*c_{2n}\langle 1-2m\vert 1-2k\rangle\langle 2k\vert 2n\rangle\langle 2m\vert 2l\rangle\\ &\langle 1-2l\vert 1-2n\rangle+ 
c_{1-2m}^*c_{1-2n} \text{sin}^2(\theta)\text{cos}^2(\chi)\langle 1-2m\vert 1-2k\rangle  \\ &\langle 1-2k\vert 1-2n\rangle \langle 2m\vert 2l\rangle\langle 2l\vert 2n\rangle \bigg ]
\end{aligned}
\end{equation}
Now using the inner product 
\begin{equation}
\begin{aligned}
C(\theta,\chi) =  \sum_{k= -\infty}^{+\infty}\bigg [ & c_{2k}^*c_{1-2k} \text{cos}^2(\theta)\text{sin}^2(\chi)+ \\ & \text{cos}(\theta)\text{sin}(\theta)\text{cos}(\chi)\text{sin}(\chi)(c_{2k}^*c_{2k} + c_{1-2k}^*c_{2k})+\\ &
c_{1-2k}^*c_{1-2k} \text{sin}^2(\theta)\text{cos}^2(\chi) \bigg ]\label{23}
\end{aligned}
\end{equation}
With Eq.\ref{6b} we can argue that $\vert c_{2k}\vert =  \vert c_{1-2k}\vert $. Also we consider there is no phase between the states $\vert 2k\rangle_1 \vert 1-2k \rangle_2 $ and $\vert 1-2k\rangle_1 \vert 2k \rangle_2$. Thus the joint probability reduces to   
\begin{align}
C(\theta,\chi) =&  \sum_{k= -\infty}^{+\infty} \vert c_{2k}\vert^2  [ \text{cos}^2(\theta)\text{sin}^2(\chi)+2\text{cos}(\theta)\text{sin}(\theta)\text{cos}(\chi)\text{sin}(\chi)+ \text{sin}^2(\theta)\text{cos}^2(\chi) ] \nonumber\\
=& \text{cos}^2(\theta-\chi). \label{23a}
\end{align}
We can have a parameter 
\begin{equation}
 E(\theta,\chi) =\frac{C(\theta,\chi)+C(\theta_{\perp},\chi_{\perp})-C(\theta_{\perp},\chi)-C(\theta,\chi_{\perp})}{C(\theta,\chi)+C(\theta_{\perp},\chi_{\perp})+C(\theta_{\perp},\chi)+C(\theta,\chi_{\perp})}
\end{equation}
and the Bell's inequality \cite{chsh} can be calculated as
\begin{equation}
B(\theta,\theta ',\chi,\chi ') = \vert E(\theta,\chi)-E(\theta,\chi ')+ E(\theta ',\chi)+E(\theta ',\chi ')\vert \leq 2 \label{12}
\end{equation}
With $\theta =0^{\circ}, \chi =22.5^{\circ} $ in Eq. \ref{23a} give a maximum Bell's inequality violation of $2\sqrt{2}$. One can check the Bell inequality for a two photon state in even/odd OAM basis with the setup given in Fig. \ref{fg.5}.  Thus  Ekert protocol \cite{Ekert} can be implemented by choosing proper measurement settings. 

\subsection{Superdense coding}

In super dense coding, Alice and Bob share entangled pair of photons. Alice encodes two bits of classical information by applying unitary operation on her entangled photon changing the combined state from one Bell state to another. Thus, by acting on one particle she can encode 2 bits of information. Alice sends her entangled particle to Bob and he does a complete Bell state analysis on both the photons, which discriminate all the Bell states. But, the efficiency of the experimental Bell state analysis is very low. In polarization entanglement, all the Bell states has not been distinguished efficiently and deterministically. At the same time if we use entanglement in another degree of freedom,  one can distinguish all the Bell states. This is known as hyper-entanglement assisted Bell state analysis (HBSA) \cite{wei, Barreiro2008}.

 We describe hyper-entanglement assisted super dense coding protocol with even/odd OAM entanglement. Consider a two photon state which is entangled both in polarization and OAM. 
 \begin{equation}
 \vert \Psi\rangle_{12}  = \vert \beta^p\rangle\otimes\vert \Psi^o\rangle
 \end{equation}
where $\vert \beta^p\rangle$ is one of the polarization Bell states and 
\begin{equation}
\begin{aligned}
\vert \Psi^o\rangle =& \sum_{k=-\infty}^{+\infty} c_{k} (\vert 2k\rangle_1 \vert 1-2k \rangle_2 + 
 \vert 2k+1\rangle_1 \vert -2k \rangle_2)\\
 \equiv & \frac{1}{\sqrt{2}}\left(\vert E\rangle_1 \vert O \rangle_2 +\vert O\rangle_1 \vert E \rangle_2 \right)
\end{aligned}
\end{equation}
 Alice encodes her two bits of information in the polarization Bell states. The final states will be 
\begin{equation}
\begin{aligned}
\vert \Psi^p\rangle^{\pm}\otimes\vert \Psi^o\rangle =& \frac{1}{{2}}\left(\vert H\rangle_1 \vert V \rangle_2 \pm \vert V\rangle_1 \vert H \rangle_2 \right)\otimes\left(\vert E\rangle_1 \vert O \rangle_2 +\vert O\rangle_1 \vert E \rangle_2 \right),\\
\vert \Phi^p\rangle^{\pm}\otimes\vert \Psi^o\rangle =& \frac{1}{{2}}\left(\vert H\rangle_1 \vert H \rangle_2 \pm \vert V\rangle_1 \vert V \rangle_2 \right)\otimes\left(\vert E\rangle_1 \vert O \rangle_2 +\vert O\rangle_1 \vert E \rangle_2 \right).
\end{aligned}
\end{equation}
Expanding 
 \begin{align}
\nonumber \vert \Psi^p\rangle^{\pm}\otimes\vert \Psi^o\rangle = \frac{1}{{2}}&(\vert H\rangle_1\vert E\rangle_1 \vert V \rangle_2\vert O \rangle_2+\vert H\rangle_1\vert O\rangle_1 \vert V \rangle_2\vert E \rangle_2 \pm\\& \vert V\rangle_1\vert E\rangle_1 \vert H \rangle_2\vert O \rangle_2\pm \vert V\rangle_1\vert O\rangle_1 \vert H \rangle_2\vert E \rangle_2 )\label{18a}\\
\nonumber \vert \Phi^p\rangle^{\pm}\otimes\vert \Psi^o\rangle = \frac{1}{{2}}&(\vert H\rangle_1\vert E\rangle_1 \vert H \rangle_2\vert O \rangle_2+\vert H\rangle_1\vert O\rangle_1 \vert H \rangle_2\vert E \rangle_2 \pm\\& \vert V\rangle_1\vert E\rangle_1 \vert V \rangle_2\vert O \rangle_2\pm \vert V\rangle_1\vert O\rangle_1 \vert V \rangle_2\vert E \rangle_2 )\label{18}
\end{align}

Single particle two qubit Bell states are defined as
\begin{equation}
\begin{aligned}
 \psi^{\pm} =&  \frac{1}{\sqrt{2}}\left( \vert H\rangle\vert E\rangle \pm \vert V\rangle\vert O\rangle\right),\\
  \phi^{\pm} =& \frac{1}{\sqrt{2}}\left( \vert H\rangle\vert O\rangle \pm \vert V\rangle\vert E\rangle\right).\label{19}
\end{aligned}
\end{equation}
 Using Eq.\ref{19} in  Eq.\ref{18a} and Eq.\ref{18} we get 
 
\begin{align}
\nonumber \vert \Psi^p\rangle^{\pm}\otimes\vert \Psi^o\rangle = \frac{1}{{4}}&((\psi^{+}_1+\psi^{-}_1) (\psi^{+}_2-\psi^{-}_2)+(\phi^{+}_1+\phi^{-}_1) (\phi^{+}_2-\phi^{-}_2) \pm\\&(\phi^{+}_1-\phi^{-}_1) (\phi^{+}_2+\phi^{-}_2)\pm(\psi^{+}_1-\psi^{-}_1) (\psi^{+}_2+\psi^{-}_2) )\label{20}\\
\nonumber \vert \Phi^p\rangle^{\pm}\otimes\vert \Psi^o\rangle = \frac{1}{{2}}&((\psi^{+}_1+\psi^{-}_1)(\phi^{+}_2+\phi^{-}_2)+(\phi^{+}_1+\phi^{-}_1) (\psi^{+}_2+\psi^{-}_2) \pm\\& (\phi^{+}_1-\phi^{-}_1) (\psi^{+}_2-\psi^{-}_2)\pm (\psi^{+}_1-\psi^{-}_1) (\phi^{+}_2-\phi^{-}_2)  )\label{21}
\end{align} 
 
giving 
\begin{equation}
\begin{aligned}
 \vert \Psi^p\rangle^{+}\otimes\vert \Psi^o\rangle = &\frac{1}{{2}}(\psi^{+}_1\psi^{+}_2-\psi^{-}_1\psi^{-}_2+ \phi^{+}_1\phi^{+}_2-\phi^{-}_1\phi^{-}_2)\label{22}\\
 \vert \Psi^p\rangle^{-}\otimes\vert \Psi^o\rangle = &\frac{1}{{2}}(\psi^{-}_1\psi^{+}_2-\psi^{+}_1\psi^{-}_2+ \phi^{-}_1\phi^{+}_2-\phi^{+}_1\phi^{-}_2)\\
  \vert \Phi^p\rangle^{+}\otimes\vert \Psi^o\rangle = &\frac{1}{{2}}(\psi^{+}_1\phi^{+}_2-\psi^{-}_1\phi^{-}_2+ \phi^{+}_1\psi^{+}_2-\phi^{-}_1\psi^{-}_2)\\
   \vert \Phi^p\rangle^{-}\otimes\vert \Psi^o\rangle = &\frac{1}{{2}}(\psi^{-}_1\phi^{+}_2-\psi^{+}_1\phi^{-}_2+ \phi^{-}_1\psi^{+}_2-\phi^{+}_1\psi^{-}_2)
\end{aligned} 
\end{equation}
 
\begin{figure}[h]
\centering
\includegraphics[scale=.4]{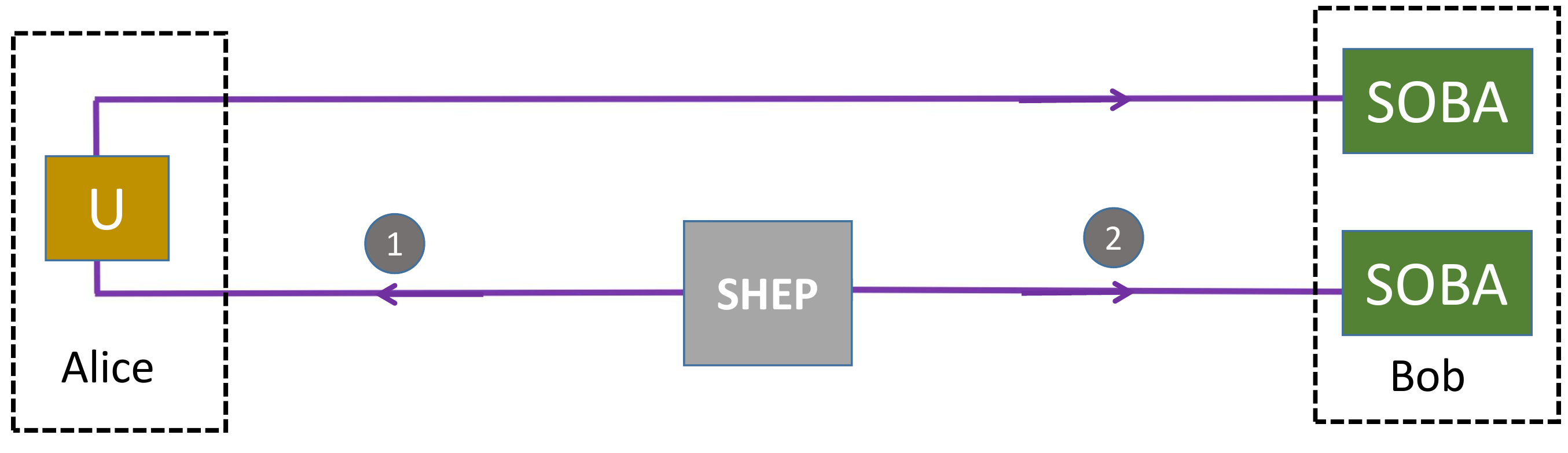}
 \caption{Setup for hyper entanglement assisted super sense coding. SHEP - source of hyper-entangled photons. }\label{fg.6}
\end{figure} 
 
 The individual single photon spin-orbit Bell state can be distinguished using SOBA which is given in Fig. \ref{fg.4}. Thus one can achieve efficient dense coding using hyper-entanglement assisted Bell state analysis. This Bell state analysis is interesting since it can be done with LOCC. A schematic for super dense coding with hyper entanglement is given in Fig. \ref{fg.6}. 
 
\section{Conclusion} 
 We have described the possibility of using even/odd OAM states for quantum information. We formulate appropriate measurement system for the even/odd OAM state based quantum information. Since even-odd OAM entanglement can be implemented like two qubit polarization entanglement, we describe the measurement and violation of Bell's inequality for such states. We have described hyper-entanglement and hybrid entanglement with OAM and polarization degrees of freedom. We have proposed an experimental scheme for spin orbit Bell state analysis to distinguish all the spin-orbit Bell states. This is applied in hyper entanglement assisted polarization Bell state analysis for efficient dense coding.





\end{document}